\documentclass{article}
\usepackage{fullpage}
\usepackage{amsmath}
\usepackage{amssymb}

\input{tcilatex}
\include{bibliography}

\begin{document}

\title{Stable Isotropic Cosmological Singularities in Quadratic Gravity}
\author{John D. Barrow and Jonathan Middleton \\
DAMTP, Centre for Mathematical Sciences\\
Cambridge University, Wilberforce Road,\\
Cambridge CB3 0WA, UK}
\maketitle

\begin{abstract}
We show that, in quadratic lagrangian theories of gravity, isotropic
cosmological singularities are stable to the presence of small scalar,
vector and tensor inhomogeneities. Unlike in general relativity, a
particular exact isotropic solution is shown to be the stable attractor on
approach to the initial cosmological singularity. This solution is also
known to act as an attractor in Bianchi universes of types I, II and IX, and
the results of this paper reinforce the hypothesis that small inhomogeneous
and anisotropic perturbations of this attractor form part of the general
cosmological solution to the field equations of quadratic gravity.
Implications for the existence of a 'gravitational entropy' are also
discussed.
\end{abstract}

\section{\protect\bigskip Introduction}

There has been considerable interest in the stability properties of
isotropic initial singularities in general relativistic cosmologies. At
first, this question was bound up with the issue of whether the initial
singularity itself was a robust prediction of general relativistic
cosmologies in the presence of typical fluids, like dust and radiation, or
whether it was merely an aberration of poor coordinate choice or high
symmetry. This question was answered unambiguously by the proofs of the
first singularity theorems which approached the problem by means of the
geodesic equations and the causal structure of space-time instead of the
Einstein equations \cite{pen}. However, there still remained the question of
whether any initial singularity was likely to be isotropic. The discovery in
1967 of the high level of temperature anisotropy in the microwave background
made this question one of central importance for observational cosmology 
\cite{cmb}. It turned out that the general behaviour of cosmological
singularities dominated by matter or radiation was highly anisotropic \cite%
{misner, bkl}, even chaotic \cite{jbchaos}, and any attempt to explain the
existence and rotation of galaxies by primordial turbulence required the
initial singularity to be highly anisotropic \cite{jbturb}, while attempts
to dissipate strong initial \ anisotropies in the spirit of the chaotic
cosmology programme \cite{misnercc} had to face the problem of entropy
over-production in the very early universe \cite{bmatz}. However, it was
possible for singularities dominated by stiff matter to be close to isotropy
and 'quiescent' because the energy density of this extreme form of matter,
characteristic of a massless scalar field, diverges as fast as the most
rapidly diverging anisotropy modes on approach to a cosmological singularity 
\cite{jbquies, stiff}. More recently, the advent of inflation \cite{guth} as
a potential explanation for the observed isotropy of the visible part of the
universe removed the need to appeal to a globally isotropic singularity in
order to explain astronomical observations, or for anisotropy to be
dissipated \cite{nohair}. It is sufficient to explain the existence of local
isotropy on the scale of the observable universe and this is achieved by
natural means in the chaotic inflationary scenario \cite{linde}. Despite the
success of this approach, there remain a number of questions whose answers
depend upon the stability properties of an initial isotropic cosmological
singularity. A sufficiently isotropic and homogeneous horizon-sized patch is
needed somewhere in order for inflation to occur, and attempts to understand
the quantum nature of gravitation will need to understand whether a physical
cosmological singularity generically occurs and, if so, what its most likely
properties are.

Penrose \cite{pen2} has proposed that initial cosmological singularities are
not simply signals that the underlying gravity theory has broken down, but
play a crucial role in defining the arrow of time. Initial singularities are
thereby required to be highly isotropic -- so possessing low gravitational
'entropy', perhaps identified with the Weyl curvature -- but any final
singularity in a collapsing closed universe would necessarily be strongly
anisotropic, with high 'gravitational entropy'. Since such isotropic
singularities are special in general relativity \cite{stab}, it is necessary
to introduce this gravitational entropy hypothesis as a boundary condition
in order to justify choosing a highly isotropic initial singularity.

Other boundary conditions can also be chosen which make the present level of
expansion isotropy understandable even if the initial state is not
isotropic. Barrow \cite{pep} showed that the natural requirement that all
energy densities (including that in anisotropising gravitational waves) are
smaller than, or of order, the Planck density at the Planck time leads to a
general requirement that the microwave background anisotropy be no larger
than $O(10^{-4})$ today. Typically, inflation scenarios assume that
arbitrarily large energy densities do not arise \cite{linde}.

Here, we want to explore the stability of isotropic singularities in
higher-order generalisations of general relativity, in which the
Einstein-Hilbert lagrangian is supplemented by the addition of the quadratic
scalar, $R$, and Ricci, $R_{ab}R^{ab}$, curvature invariants. Since initial
cosmological singularities are expected to involve infinities in one of more
of the curvature invariants of the space-time, we expect that the addition
of higher-order terms to the lagrangian would produce a new dominant
behaviour to such singularities. We would expect the higher-order terms to
control the dynamics as the singularity is approached, and to determine the
stability properties of any isotropic special solution in that limit. In
this paper we are going to consider the contributions of the quadratic Ricci
terms, $R_{ab}R^{ab}$, to the lagrangian. The quadratic scalar, $R^{2}$,
contributions are conformally equivalent to the presence of a
self-interacting scalar field and are understood. In a subsequent paper, we
will study how our results may be modified by complicated effects that arise
at general order, when $(R_{ab}R^{ab})^{n}$ terms are added to the
lagrangian.

\ Specifically, we will investigate the stability of an initial isotropic
zero-curvature Friedmann singularity to small inhomogeneous scalar, vector,
and tensor perturbations in generalisations of general relativity which
contain quadratic Ricci and scalar curvature contributions to the
gravitational lagrangian of general relativity. We will show that there
exists a particular isotropic solution of the pure quadratic Ricci theory
which is a stable attractor as $t\rightarrow 0$ for small inhomogeneous
perturbations of the metric when the gravitation theory created by the
addition of quadratic Ricci terms to the general relativity lagrangian.
Previous studies have shown this particular isotropic solution to be a
stable $t\rightarrow 0$ attractor for spatially homogeneous vacuum universes
of Bianchi types $I,II,$ \cite{bher1} and $IX$ \cite{cotIX}. Our results
lend further weight to the conjecture of Barrow and Hervik \cite{bher1} that
this isotropic vacuum solution of the pure quadratic Ricci theory
characterises part of the general cosmological solution in these particular
higher-order generalisations of general relativity. This behaviour is
completely different to that found in general relativity, where isotropy is
unstable in the $t\rightarrow 0$ limit of an initial cosmological
singularity \cite{misner, bkl}.

\section{A Special Solution}

Consider a quadratic gravity theory with action

\begin{equation*}
S=\dint d^{4}x\sqrt{-g}\left[ \frac{1}{2}(-2\Lambda
+R+AR^{2}+BR_{ab}R^{ab})+L_{m}\right] .
\end{equation*}%
The field equations obtained by varying this action with respect to the
metric are \cite{deser, bher1, bher2}:

\begin{equation}
G_{b}^{a}+AP^{(1)a}\!_{b}+BP^{(2)a}\!_{b}=T_{b}^{a}-\Lambda g_{b}^{a},
\end{equation}%
where 
\begin{eqnarray*}
P^{(1)a}\!_{b} &=&-\frac{1}{2}R^{2}g_{b}^{a}+2RR_{b}^{a}+2g_{b}^{a}\Box
R-2R_{;}\,\!^{a}\,\!_{b}\,, \\
P^{(2)a}\!_{b} &=&-\frac{1}{2}R^{cd}R_{cd}g_{b}^{a}+\Box R_{b}^{a}+\frac{1}{2%
}g_{b}^{a}\Box R-R_{;}\,\!^{a}\,\!_{b}-2R^{a}\,\!_{cdb}R^{cd}\,,
\end{eqnarray*}%
where $G_{ab}=R_{ab}-\frac{1}{2}Rg_{ab}$ is the usual Einstein tensor, $%
T_{b}^{a}$ is the energy-momentum tensor of the matter sources, and $\Lambda 
$ is the cosmological constant\footnote{%
For a different covariant formulation see ref. \cite{maar}.}.

We take $\Lambda =0$ and consider perturbations about the spatially flat,
homogeneous and isotropic FRW spacetime, 
\begin{equation}
ds^{2}=-dt^{2}+a^{2}(t)(dx^{2}+dy^{2}+dz^{2}),
\end{equation}%
with scale factor $a(t)$and associated Hubble expansion rate $H\equiv \dot{a}%
/a$, for which the background equations are: 
\begin{eqnarray}
\frac{1}{3}\mu  &=&H^{2}+2(B+3A)\left( 2H\ddot{H}-\dot{H}^{2}+6H^{2}\dot{H}%
\right) , \\
-\frac{1}{2}(\mu +p) &=&\dot{H}+2(B+3A)\left( \dddot{H}+3H\ddot{H}+6\dot{H}%
^{2}\right) ,
\end{eqnarray}%
where we have defined a density, $\mu =-T_{0}^{0},$ and pressure $p=\frac{1}{%
3}T_{\gamma }^{\gamma }$.

For radiation obeying the equation of state $p=\frac{1}{3}\mu $, there
exists a special exact solution of these equations with \cite{special, bher2}

\begin{equation*}
a(t)=t^{\frac{1}{2}}.
\end{equation*}

Importantly, this is also an exact \textit{vacuum} solution ($\mu =p=0$) of
the purely quadratic theory\footnote{%
Barrow and Hervik \cite{bher1} showed that this is also true for the case of
Friedmann models with non-zero curvature, in which case the curved Friedmann
radiation solution is an exact solution of the full equations in the
presence of radiation and also of the purely quadratic vacuum equations $%
(B+3A\rightarrow \infty ).$} ($B+3A\rightarrow \infty $), i.e. it solves 
\begin{equation}
2H\ddot{H}-\dot{H}^{2}+6H^{2}\dot{H}=0.
\end{equation}

Integrating this equation, we have

\begin{equation*}
\frac{\dot{H}^{2}}{H}=\frac{M^{2}}{a^{6}},
\end{equation*}

with $M$ constant. Since $\dot{H}=aH\frac{dH}{da}$ we can take the square
root and integrate to obtain 
\begin{equation*}
\frac{2}{3}H^{\frac{3}{2}}=D\pm \frac{M}{3a^{3}},
\end{equation*}%
with $D$ constant. Hence, we can see that $a\rightarrow t^{\frac{1}{2}}$ as $%
t\rightarrow 0$ and $H^{2}\rightarrow const$ $+O(a^{-3})$ to leading order
as $a\rightarrow \infty .$ So the purely quadratic vacuum solution mimics
radiation early on, then dust plus a cosmological constant at late times.

Clifton and Barrow \cite{Clifton1} found that for a universe filled with a
fluid with equation of state $p=\omega \mu $ in a theory where the
lagrangian depends only on the Ricci-squared term like $(R_{ab}R^{ab})^{n}$,
there is an isotropic fluid-filled FRW solution with

\begin{equation*}
a(t)=t^{\frac{4n}{3(\omega +1)}},
\end{equation*}%
and also that there exists an isotropic vacuum solution with

\begin{eqnarray}
a &=&t^{1/2}:n=1,  \label{n1} \\
a &=&t^{S}:n\neq 1,  \label{n2}
\end{eqnarray}%
where \cite{Clifton1}

\begin{equation*}
S\equiv \frac{3(1-3n+4n^{2})\pm \sqrt{3(48n^{4}-40n^{3}-5n^{2}+10n-1)}}{%
6(1-n)}.
\end{equation*}

In particular, for the case of $n=1$ and $\omega \leq 1$, we see that $\frac{%
4n}{3(\omega +1)}\geq \frac{2}{3}>\frac{1}{2}$, so the special vacuum
solution will dominate over the fluid-filled effects in the field equations
as $t\rightarrow 0$.

The special isotropic vacuum solution (\ref{n1}) is of special interest.
Previous studies have shown that it is an attractor for solutions of Bianchi
types I and II as $t\rightarrow 0$, \cite{bher1}. It is also a stable
attractor in Bianchi type IX 'Mixmaster' universes as $t\rightarrow 0$ \cite%
{cotIX}. This is in complete contrast to the situation that exists in
general relativity ($A=B=0$) where the shear and curvature anisotropy terms
dominate the dynamics as the singularity is approached at $t=0,$ \cite%
{misner, bkl}; in the case of Bianchi type IX, that dynamical approach is
chaotic \cite{jbchaos}. As a result an isotropic initial singularity
constitutes a set of zero measure in the initial data space. In quadratic
gravity, the situation is completely different, the results in Bianchi types
I, II, and IX suggest that the special isotropic solution may be an
attractor for the time evolution of a part of the general solution of the
cosmological evolution equations. In the sections to follow we will show
that this special vacuum solution is also stable against the effects of
small inhomogeneous perturbations (scalar, vector, and tensor) as $%
t\rightarrow 0$.

\section{Inhomogeneous Perturbations}

The general perturbation of the FRW metric may be written as 
\begin{equation}
ds^{2}=-a^{2}(1+2\alpha )d\eta ^{2}-a^{2}\tilde{B}_{\alpha }d\eta dx^{\alpha
}+a^{2}(\delta _{\alpha \beta }+\tilde{C}_{\alpha \beta })dx^{\alpha
}dx^{\beta },
\end{equation}%
where $\eta $ is a conformal time coordinate that is related to the comoving
proper time, $t$, by $dt=ad\eta $. We can decompose the perturbation
variables into their \textquotedblleft scalar\textquotedblright ,
\textquotedblleft vector\textquotedblright\ and \textquotedblleft
tensor\textquotedblright\ parts, as follows\footnote{%
Note that there is a factor of 2 difference in the $d\eta dx^{\alpha }$
component from the metrics given by Noh and Hwang in \cite{Noh:vort} and 
\cite{Noh:scalar}. Defining $g_{0\alpha }=-a^{2}(\beta _{,\alpha }+B_{\alpha
})$ is consitent with their equations and appendix quantities.}:%
\begin{eqnarray*}
\tilde{B}_{\alpha } &=&2\beta _{,\alpha }+2B_{\alpha } \\
\tilde{C}_{\alpha \beta } &=&2\phi \delta _{\alpha \beta }+2\gamma _{,\alpha
\beta }+2C_{(\alpha ,\beta )}+2C_{\alpha \beta }.
\end{eqnarray*}

There are four scalar perturbation variables, $\alpha ,\beta ,\phi $ and $%
\gamma $, two vector variables, $B_{\alpha }$ and $C_{\alpha },$ and one
tensor perturbation, $C_{\alpha \beta }$. The quantities $B_{\alpha }$ and $%
C_{\alpha }$ are divergence-free, i.e. $B^{\alpha }\!_{,\alpha }\equiv
0\equiv C^{\alpha }\!_{,\alpha }$, and $C_{\alpha \beta }$ is transverse and
tracefree, i.e. $C^{\alpha }\!_{\beta ,\alpha }=0=C^{\alpha }\!_{\alpha }$.
These three types of perturbation evolve independently of each other at
linear order. The linearised equations governing the perturbations in the
general quadratic gravity are presented in a series of papers by Noh and
Hwang \cite{Noh:gw, Noh:vort, Noh:scalar}.

For the gravitational-wave (tensor) perturbation modes in the general
quadratic gravity, the equation for the perturbation is (\cite{Noh:gw}): 
\begin{eqnarray*}
\delta T_{\beta }^{\alpha } &=&D_{\beta }^{\alpha }+2A\left( RD_{\beta
}^{\alpha }+\dot{R}\dot{C}_{\beta }^{\alpha }\right) -B\biggl\{\ddot{D}%
_{\beta }^{\alpha }+3H\dot{D}_{\beta }^{\alpha }-\frac{\Delta }{a^{2}}\left(
D_{\beta }^{\alpha }+4\dot{H}C_{\beta }^{\alpha }\right)  \\
&&-6\biggl[\left( \dot{H}+H^{2}\right) D_{\beta }^{\alpha }+\left( \ddot{H}+H%
\dot{H}\right) \dot{C}_{\beta }^{\alpha }\biggr]\biggr\}
\end{eqnarray*}%
where 
\begin{equation*}
D_{\beta }^{\alpha }\equiv \ddot{C}_{\beta }^{\alpha }+3H\dot{C}_{\beta
}^{\alpha }-\frac{\Delta }{a^{2}}C_{\beta }^{\alpha }.
\end{equation*}

For the vortical (vector) perturbations, following Noh and Hwang \cite%
{Noh:vort}, we set $B_{\alpha }(t,\mathbf{x})\equiv b(t)Y_{\alpha }(\mathbf{x%
})$ , with $\Delta Y_{\alpha }\equiv -k^{2}Y_{\alpha }$ and use the
gauge-invariant variable $\Psi Y_{\alpha }\equiv B_{\alpha }+a\dot{C}%
_{\alpha }$. In the gauge where $C_{\alpha }\equiv 0$, the equations are 
\begin{equation*}
\frac{k^{2}}{2a^{2}}\left\{ (1+2AR)\Psi -B\left[ \ddot{\Psi}+H\dot{\Psi}%
-\left( \frac{2}{3}R-\frac{k^{2}}{a^{2}}\right) \Psi \right] \right\} =(\mu
+p)v_{\omega }
\end{equation*}%
and the conservation of angular momentum transparently holds: 
\begin{equation*}
a^{4}(\mu +p)v_{\omega }=\Omega
\end{equation*}%
with $\Omega $ constant.

Finally, for the scalar density perturbations, we introduce the combinations 
$\chi \equiv a(\beta +a\dot{\gamma})$ and $\kappa \equiv 3(H\alpha -\dot{\phi%
})-\frac{\Delta \chi }{a^{2}}$. Then we have the following set of equations (%
\cite{Noh:scalar}):

Energy constraint: 
\begin{eqnarray*}
\delta T_{0}^{0} &=&2\left( \frac{\Delta \phi }{a^{2}}+H\kappa \right) +2A%
\biggl\{2R\left( \frac{\Delta \phi }{a^{2}}+H\kappa \right) +\dot{R}\left(
\kappa +3H\alpha \right) -3H\delta \dot{R}+\left[ 3(\dot{H}+H^{2})+\frac{%
\Delta }{a^{2}}\right] \delta R\biggr\} \\
&&-B\biggl\{2\left( H\kappa +\frac{\Delta \phi }{a^{2}}\right)
^{..}+6H\left( H\kappa +\frac{\Delta \phi }{a^{2}}\right) ^{.}+\left[ 4(\dot{%
H}-6H^{2})-\frac{2\Delta }{a^{2}}\right] \left( H\kappa +\frac{\Delta \phi }{%
a^{2}}\right) +3H\delta \dot{R} \\
&&-\left( 3H^{2}+\frac{\Delta }{a^{2}}\right) \delta R+6H\dot{H}\dot{\alpha}%
+6\left( -H\ddot{H}+2\dot{H}^{2}-9H^{2}\dot{H}\right) \alpha  \\
&&-\left( 6\dot{H}+5H^{2}\right) ^{.}\kappa -\frac{8H}{3}\frac{\Delta }{a^{2}%
}\left( \kappa +\frac{\Delta \chi }{a^{2}}\right) \biggr\}.
\end{eqnarray*}

Momentum constraint: 
\begin{eqnarray*}
T^0_\alpha &=& \frac{2}{3a}\nabla_\alpha\Biggl[-\kappa -\frac{\Delta \chi}{%
a^2}+A\left(-2R\kappa -2R\frac{\Delta \chi}{a^2} -3\dot{R}\alpha+3\delta\dot{%
R}-3H\delta R\right)  \notag \\
&& -B\biggl\{-\left(\kappa + \frac{\Delta \chi}{a^2}\right)^{..}-3H\left(%
\kappa + \frac{\Delta \chi}{a^2}\right)^{.} + \left(12H^2 +\frac{\Delta}{a^2}
\right)\left(\kappa + \frac{\Delta \chi}{a^2}\right)-\frac{3\delta \dot{R}}{2%
}+\frac{3H \delta R}{2}  \notag \\
&& -3\dot{H}\dot{\alpha}+\frac{3}{2}\left(2\dot{H}+9H^2 \right)^{.}\alpha+%
\frac{3\Delta}{a^2}\left[2H\phi+\left(\dot{H}-2H^2\right) \chi \right]%
\biggr\}\Biggr]
\end{eqnarray*}

Tracefree propagation: 
\begin{eqnarray*}
\delta T_{\beta }^{\alpha }-\frac{1}{3}\delta T_{\gamma }^{\gamma }\delta
_{\beta }^{\alpha } &=&\frac{1}{a^{2}}\left( \nabla ^{\alpha }\nabla _{\beta
}-\frac{1}{3}\delta _{\beta }^{\alpha }\Delta \right) \Biggl[\dot{\chi}%
+H\chi -\phi -\alpha +2A\left( R\dot{\chi}+(HR+\dot{R})\chi -\delta R-\phi
-\alpha \right)  \\
&&-B\biggl\{(\dot{\chi}+H\chi -\phi -\alpha )^{..}-H(\dot{\chi}+H\chi -\phi
-\alpha )^{.} \\
&&-\left[ 8(\dot{H}+H^{2})+\frac{\Delta }{a^{2}}\right] (\dot{\chi}+H\chi
-\phi -\alpha ) \\
&&+\delta R-4\dot{H}\phi -6\left( \dot{H}+H^{2}\right) ^{.}\chi +\frac{8H}{3}%
\left( \kappa +\frac{\Delta \chi }{a^{2}}\right) \biggr\}\Biggr].
\end{eqnarray*}

Raychaudhuri equation: 
\begin{eqnarray*}
\delta T_{\gamma }^{\gamma }-\delta T_{0}^{0} &=&2\dot{\kappa}+4H\kappa
+2\left( 3\dot{H}+\frac{\Delta }{a^{2}}\right) \alpha +2A\biggl\{2R\dot{%
\kappa}+\left( 4HR+\dot{R}\right) \kappa +3\dot{R}\dot{\alpha} \\
&&+\left[ 6\ddot{R}+3H\dot{R}+2R\left( 3\dot{H}+\frac{\Delta }{a^{2}}\right) %
\right] \alpha -3\delta \ddot{R}-3H\delta \dot{R}+\left( 6H^{2}+\frac{\Delta 
}{a^{2}}\right) \delta R\biggr\} \\
&&-B\biggl\{-4\left( H\kappa +\frac{\Delta \phi }{a^{2}}\right)
^{..}-12H\left( H\kappa +\frac{\Delta \phi }{a^{2}}\right) ^{.} \\
&&+\left[ -8\left( \dot{H}-6H^{2}\right) +\frac{4\Delta }{a^{2}}\right]
\left( H\kappa +\frac{\Delta \phi }{a^{2}}\right) +2\delta \ddot{R} \\
&&+6H^{2}\delta R-28\dot{H}\kappa -6\left( 2\dot{H}+5H^{2}\right) \dot{\alpha%
}-12\left[ \left( 2\dot{H}+5H^{2}\right) ^{..}+3H^{2}\dot{H}\right] \alpha \\
&&+\frac{16H}{3}\frac{\Delta }{a^{2}}\left( \kappa +\frac{\Delta \chi }{a^{2}%
}\right) \biggr\}
\end{eqnarray*}

Trace equation: 
\begin{equation*}
\delta T = - \delta R -2(3A+B)\left[\delta \ddot{R}+3H\delta \dot{R}-\frac{%
\Delta}{a^2}\delta R - \dot{R}(\kappa+\dot{\alpha})-\left(2\ddot{R}+3H\dot{R}%
\right)\alpha \right]
\end{equation*}

and 
\begin{equation*}
\delta R = -2\left[\dot{\kappa} +4H\kappa +3\dot{H}\alpha +\frac{\Delta}{a^2}%
(2\phi +\alpha)\right]
\end{equation*}

Note that no choice of gauge has yet been made for the scalar perturbations;
we will ultimately want to work in the gauge $\delta R \equiv 0$, so that we
are considering perturbations in flat space.

We are primarily interested in the effects of the $R_{ab}R^{ab}$ term and so
henceforth we set $A=0$\footnote{%
Note that for our special solution, eq. (\ref{n1}) with $a(t)=t^{\frac{1}{2}}
$, we have $R=0$ to background order. Consequently, the perturbed parts of
the gravitational wave and vorticity equations are independent of $A$. This
is also true for the scalar perturbations when we choose the uniform
curvature gauge, i.e. $\delta R\equiv 0$. So our results are also valid for
the general quadratic gravity.}. We consider the three perturbation modes in
turn in order to determine whether the isotropic and homogeneous vacuum
solution (\ref{n1}) is stable against their effects as $t\rightarrow 0$.

\section{Gravitational-wave perturbations}

The metric for the gravitational-wave perturbations takes the form 
\begin{equation*}
ds^{2}=-dt^{2}+a^{2}(\delta _{\alpha \beta }+2C_{\alpha \beta })dx^{\alpha
}dx^{\beta },
\end{equation*}%
\begin{equation}
C_{\alpha }^{\alpha }=0=C_{\beta ,\alpha }^{\alpha }  \label{con}
\end{equation}%
where $C_{\alpha \beta }=C_{\alpha \beta }(\mathbf{x},t)$ and is
gauge-invariant. For convenience, from now on we drop the indices on $%
C_{\alpha \beta }$ . The remaining perturbation equations in vacuum are: 
\begin{eqnarray*}
0 &=&D-B\left\{ \ddot{D}+3H\dot{D}-\frac{\Delta }{a^{2}}(D+4\dot{H}C)-6\left[
(\dot{H}+H^{2})D+(\dot{H}+H^{2}\dot{)}\dot{C}\right] \right\}  \\
D &\equiv &\ddot{C}+3H\dot{C}-\frac{\Delta C}{a^{2}},
\end{eqnarray*}%
where $D\equiv D_{\beta }^{\alpha }$ and $C\equiv C_{\beta }^{\alpha }$.

Substituting $H=1/2t$, and taking the $B\rightarrow \infty $ limit, keeping
only $\left\{ {}\right\} =0$, this reduces to 
\begin{eqnarray}
\ddot{D}+\frac{3}{2t}\dot{D}+\frac{3}{2t^{2}}D-\frac{3}{t^{3}}\dot{C} &=&%
\frac{\Delta }{t}\left( D-\frac{2}{t^{2}}C\right) ,  \label{m1} \\
\ddot{C}+\frac{3}{2t}\dot{C}-\frac{\Delta C}{t} &=&D.  \label{m2}
\end{eqnarray}

Substituting for $D$ using (\ref{m2}), we have the single 4th-order
evolution equation for the metric perturbation $C$: 
\begin{equation}
0=\ddddot{C}+\frac{3}{t}\dddot{C}+\frac{3}{4t^{2}}\ddot{C}-\frac{2\Delta 
\ddot{C}}{t}-\frac{\Delta \dot{C}}{t^{2}}+\frac{\Delta \Delta C}{t^{2}}.
\label{4th}
\end{equation}

\subsection{Large scales}

In the long-wavelength limit, the last three terms in (\ref{4th}) can be
neglected and we have 
\begin{equation*}
0=\ddddot{C}+\frac{3}{t}\dddot{C}+\frac{3}{4t^{2}}\ddot{C},
\end{equation*}%
and so the solution is a linear sum of power-laws with 
\begin{eqnarray*}
C &\propto &t^{q}, \\
0 &=&q(q-1)(2q-1)(2q-3),
\end{eqnarray*}%
hence

\begin{equation*}
C=\varepsilon +\lambda t+\nu t^{\frac{1}{2}}+\zeta t^{\frac{3}{2}},
\end{equation*}%
where $\varepsilon ,\lambda ,\nu ,\varsigma $ are independent arbitrary
tensor functions of $\mathbf{x}$ satisfying the same constraints as $C,$
given by (\ref{con}).

Note that as $t\rightarrow 0,$ the perturbation $C\rightarrow const$ and the 
$a=t^{\frac{1}{2}}$ solution is stable against gravitational-wave modes. The
leading-order terms in the metric perturbation expansion are
therefore(restoring the tensor indices) : 
\begin{equation*}
ds^{2}=-dt^{2}+t(\delta _{\alpha \beta }+2\varepsilon _{\alpha \beta }+2\nu
_{\alpha \beta }t^{\frac{1}{2}}+2\lambda _{\alpha \beta }t+2\zeta _{\alpha
\beta }t^{\frac{3}{2}})dx^{\alpha }dx^{\beta },
\end{equation*}%
where the $\varepsilon _{\alpha \beta }(\mathbf{x}),\nu _{\alpha \beta }(%
\mathbf{x}),\lambda _{\alpha \beta }(\mathbf{x}),\zeta _{\alpha \beta }(%
\mathbf{x})$ are arbitrary traceless symmetric 3x3 tensors subject only to (%
\ref{con}) and coordinate transformations and the constraint equations.

\subsection{Small scales}

If we look for a separable solution to (\ref{4th}) of the form 
\begin{equation*}
C=T(t)X(\mathbf{x})
\end{equation*}

then 
\begin{eqnarray*}
\ddddot{C}+\frac{3}{t}\dddot{C}+\frac{3}{4t^{2}}\ddot{C} &=&\frac{2\Delta 
\ddot{C}}{t}+\frac{\Delta \dot{C}}{t^{2}}-\frac{\Delta \Delta C}{t^{2}} \\
\Rightarrow \ddddot{T}X+\frac{3}{t}\dddot{T}X+\frac{3}{4t^{2}}\ddot{T}X &=&%
\frac{2\ddot{T}\Delta X}{t}+\frac{\dot{T}\Delta X}{t^{2}}-\frac{T\Delta
\Delta X}{t^{2}}.
\end{eqnarray*}%
If $\Delta X=\theta X$, where $\theta $ is constant so $\Delta \Delta
X=\theta \Delta X=\theta ^{2}X$, then 
\begin{equation}
\ddddot{T}+\frac{3}{t}\dddot{T}+\frac{3}{4t^{2}}\ddot{T}=\frac{2\ddot{T}%
\theta }{\ t}+\frac{\dot{T}\ \theta }{\ t^{2}}-\frac{\ \theta ^{2}T}{\ t^{2}}%
,  \label{small}
\end{equation}%
\begin{equation}
t^{2}\ddddot{T}+3t\dddot{T}+\left( \frac{3}{4}-2t\theta \right) \ddot{T}-%
\dot{T}\ \theta +\theta ^{2}T=0.  \label{master}
\end{equation}

If $\theta =0$, then we obtain the same results as in large-scale limit. If
we take the late-time approximation or the small-scale limit by keeping only
the terms in $\theta <0$ in (\ref{small}) then we need to solve 
\begin{equation*}
2t\ddot{T}+\dot{T}-\theta T=0,
\end{equation*}

so as $\theta $ will be negative, 
\begin{equation*}
T=C_{1}\cos \{\sqrt{-2\theta t}\}+C_{2}\sin \{\sqrt{-2\theta t}\}
\end{equation*}
which approaches a constant as $t\rightarrow 0$.

\subsection{The general separable solution}

If we change to conformal time, $t=\frac{\eta ^{2}}{4}$, and put $\theta
=-k^{2}$, then the master equation (\ref{master}) is transformed to 
\begin{equation*}
T^{\prime \prime \prime \prime }+2k^{2}T^{\prime \prime }+k^{4}T=0.
\end{equation*}%
The general solution is 
\begin{eqnarray*}
T^{\ast }(t)=A_{1}\cos [2k\sqrt{t}]+A_{2}\sin [2k\sqrt{t}]+A_{3}\sqrt{t}\cos
[2k\sqrt{t}]+A_{4}\sqrt{t}\sin [2k\sqrt{t}] &&\mbox{\, $k \neq
  0$}, \\
T=B_{1}+B_{2}\sqrt{t}+B_{3}t+B_{4}t\sqrt{t} &&\mbox{\, $k=0$}.
\end{eqnarray*}%
We note that, for $k\neq 0$, 
\begin{equation*}
T^{\ast }(t)\rightarrow A_{1}+(2kA_{2}+A_{3})\sqrt{t}+O(t)\rightarrow const
\end{equation*}%
as $t\rightarrow 0$.

Hence the full separable solution for the metric perturbation is 
\begin{eqnarray*}
C_{\alpha \beta }(t,\mathbf{x}) &=&T^{\ast }(t)X_{\alpha \beta }(\mathbf{x}),
\\
\Delta X_{\alpha \beta } &=&-k^{2}X_{\alpha \beta }.
\end{eqnarray*}

\section{Vortical Perturbations}

The metric expansion for the vorticity perturbations takes the form

\begin{equation}
ds^{2}=-a^{2}d\eta ^{2}-2a^{2}B_{\alpha }d\eta dx^{\alpha }+a^{2}\left(
\delta _{\alpha \beta }+2C_{(\alpha ,\beta )}\right) dx^{\alpha }dx^{\beta },
\end{equation}%
with $B^{\alpha }\!_{,\alpha }\equiv 0\equiv C^{\alpha }\!_{,\alpha }$.

Following Noh and Hwang \cite{Noh:vort}, we set $B_{\alpha }(t,\mathbf{x}%
)\equiv b(t)Y_{\alpha }(\mathbf{x})$ , with $\Delta Y_{\alpha }\equiv
-k^{2}Y_{\alpha },$ and use the gauge-invariant variable:

\begin{equation*}
\Psi Y_{\alpha }\equiv B_{\alpha }+a\dot{C}_{\alpha }.
\end{equation*}

In the gauge defined by $C_{\alpha }\equiv 0$, linearising about the special
vacuum solution with $H=\frac{1}{2t}$and $R=0$ we then have 
\begin{equation*}
\frac{k^{2}}{2a^{2}}\left\{ \Psi -B\left[ \ddot{\Psi}+\frac{\dot{\Psi}}{2t}+%
\frac{k^{2}}{t}\Psi \right] \right\} =(\mu +p)v_{\omega }
\end{equation*}

and the conservation of angular momentum gives: 
\begin{equation*}
a^{4}(\mu +p)v_{\omega }=\Omega 
\end{equation*}%
with $\Omega $ constant.

\subsection{Large scales}

On large scales we have 
\begin{equation}
\Psi -B\left[ \ddot{\Psi}+\frac{\dot{\Psi}}{2t}\right] =\frac{2\Omega }{%
k^{2}t}.  \label{vort}
\end{equation}

In the case where the (quadratic) $B$ term dominates on the LHS, we have 
\begin{equation*}
\Psi =-\frac{D}{B}t^{1/2}-\frac{4\Omega }{Bk^{2}}t+\Psi _{0}
\end{equation*}%
and $\Psi \rightarrow const$ as $t\rightarrow 0$ and the isotropic metric is
stable.

If we keep the general-relativistic $\Psi $ term on the LHS of (\ref{vort}),
and write $\Psi =t^{\frac{1}{4}}f$, then we have 
\begin{equation}
t^{2}\ddot{f}+t\dot{f}-\left( \frac{t^{2}}{B}+\frac{1}{16}\right) f=-\frac{%
2\Omega }{k^{2}B}t^{\frac{3}{4}}.  \label{in}
\end{equation}%
This is an inhomogeneous modified Bessel equation; so, if we set 
\begin{equation*}
y=\frac{it}{\sqrt{B}},\nu =\frac{1}{4},g=-\frac{2\Omega t^{\frac{3}{4}}}{%
Bk^{2}},
\end{equation*}%
we obtain 
\begin{equation*}
y^{2}f^{\prime \prime }+yf^{\prime }+(y^{2}-\nu ^{2})f=g(x)
\end{equation*}%
and its solution is 
\begin{eqnarray*}
f &=&C_{1}J_{\frac{1}{4}}\left( \frac{it}{\sqrt{B}}\right) +C_{2}Y_{\frac{1}{%
4}}\left( \frac{it}{\sqrt{B}}\right) +\frac{\pi }{2}Y_{\frac{1}{4}}\left( 
\frac{it}{\sqrt{B}}\right) \int \frac{it}{\sqrt{B}}J_{\frac{1}{4}}\left( 
\frac{it}{\sqrt{B}}\right) \left( \frac{-2\Omega t^{\frac{3}{4}}}{Bk^{2}}%
\right) \frac{i}{\sqrt{B}}dt \\
&&-\frac{\pi }{2}J_{\frac{1}{4}}\left( \frac{it}{\sqrt{B}}\right) \int \frac{%
it}{\sqrt{B}}Y_{\frac{1}{4}}\left( \frac{it}{\sqrt{B}}\right) \left( \frac{%
-2\Omega t^{\frac{3}{4}}}{Bk^{2}}\right) \frac{i}{\sqrt{B}}dt.
\end{eqnarray*}

Hence, on large-scales the general solution in the $C_{\alpha }\equiv 0$
gauge is 
\begin{eqnarray}
B_{\alpha } &\equiv &\Psi Y_{\alpha }=t^{\frac{1}{4}}f(t)Y_{\alpha }(\mathbf{%
x}), \\
\Delta Y_{\alpha } &\equiv &-k^{2}Y_{\alpha }.
\end{eqnarray}%
Again we note that $B_{\alpha }\rightarrow constant$ as $t\rightarrow 0$ and
the isotropic solution (\ref{n1}) is stable in that limit.

\subsection{The pure quadratic theory}

Now consider the case where the B-term dominates, but this time without
taking the long-wavelength limit, i.e. we drop just the GR term on the left
hand side. In that case the governing equation is 
\begin{equation}
\ddot{\Psi}+\frac{\dot{\Psi}}{2t}+\frac{k^{2}\Psi }{t}=-\frac{2\Omega }{%
Bk^{2}t}
\end{equation}%
and so 
\begin{equation}
\Psi =D_{1}\cosh \{2k\sqrt{t}\}+D_{2}\sinh \{2k\sqrt{t}\}-\frac{2\Omega }{%
Bk^{4}},
\end{equation}%
where $D_{1}$ and $D_{2}$ are constants.

Since $B_{\alpha }\equiv \Psi Y_{\alpha }(\mathbf{x})$ in our chosen gauge,
we again see that $B_{\alpha }\rightarrow constant$ as $t\rightarrow 0$ and (%
\ref{n1}) is stable in that limit.

\section{Scalar Perturbations}

For the scalar perturbations, the metric will take the form 
\begin{equation*}
ds^{2}=-a^{2}(1+2\alpha )d\eta ^{2}-2a^{2}\beta _{,\alpha }d\eta dx^{\alpha
}+a^{2}(\delta _{\alpha \beta }(1+2\phi )+2\gamma _{,\alpha \beta
})dx^{\alpha }dx^{\beta }.
\end{equation*}%
There are possible gauge choices that can still be made and we shall
consider two useful choices.

\subsection{Conformal Newtonian Gauge ($\protect\chi \equiv 0 $)}

The gauge choice $\chi \equiv a(\beta +a\dot{\gamma})=0$ (zero shear)
greatly simplifies the perturbation equations, since we just have two
perturbed variables to worry about. They reduce to the following set:

\noindent Energy constraint: 
\begin{eqnarray}
\delta T_{0}^{0} &=&-B\biggl\{6\frac{\dddot{\phi}}{t}+15\frac{\ddot{\phi}}{%
t^{2}}-6\frac{\dot{\phi}}{t^{3}}-4\frac{\Delta \ddot{\phi}}{t}-12\frac{%
\Delta \dot{\phi}}{t^{2}}+2\frac{\Delta \phi }{t^{3}}+2\frac{\Delta ^{2}\phi 
}{t^{2}}-3\frac{\ddot{\alpha}}{t^{2}}+\frac{3\dot{\alpha}}{2t^{3}}+\frac{%
\Delta \alpha }{t^{3}}+2\frac{\Delta ^{2}\alpha }{t^{2}}\biggr\}  \notag \\
&&+\frac{2\Delta \phi }{t}-\frac{3\dot{\phi}}{t}+\frac{3\alpha }{2t^{2}}.
\label{z1}
\end{eqnarray}%
Momentum constraint: 
\begin{equation}
T_{\alpha }^{0}=-\frac{2}{t^{\frac{1}{2}}}\nabla _{\alpha }\Biggl[B\biggl\{-2%
\dddot{\phi}-3\frac{\ddot{\phi}}{t}+6\frac{\dot{\phi}}{t^{2}}+\frac{\Delta 
\dot{\phi}}{t}-2\frac{\Delta \phi }{t^{2}}+\frac{\ddot{\alpha}}{t}-3\frac{%
\dot{\alpha}}{2t^{2}}+\frac{\Delta \dot{\alpha}}{t}-\frac{\Delta \alpha }{%
t^{2}}\biggr\}-\dot{\phi}+\frac{\alpha }{2t}\Biggr].  \label{z2}
\end{equation}%
Trace-free propagation: 
\begin{equation}
\delta T_{\beta }^{\alpha }-\frac{1}{3}\delta T_{\gamma }^{\gamma }\delta
_{\beta }^{\alpha }=-\frac{1}{t}\left( \nabla ^{\alpha }\nabla _{\beta }-%
\frac{1}{3}\delta _{\beta }^{\alpha }\Delta \right) \left[ B\biggl\{5\ddot{%
\phi}+\frac{17\dot{\phi}}{2t}-3\frac{\Delta \phi }{t}-\ddot{\alpha}-\frac{5%
\dot{\alpha}}{2t}-\frac{\Delta \alpha }{t}\biggr\}+\phi +\alpha \right] .
\label{z3}
\end{equation}%
Trace equation: 
\begin{eqnarray}
\delta T &=&-2B\biggl\{6\ddddot{\phi}+21\frac{\dddot{\phi}}{t}-6\frac{\ddot{%
\phi}}{t^{2}}+6\frac{\dot{\phi}}{t^{3}}-10\frac{\Delta \ddot{\phi}}{t}-10%
\frac{\Delta \dot{\phi}}{t^{2}}-2\frac{\Delta {\phi }}{t^{3}}+4\frac{\Delta
^{2}{\phi }}{t^{2}}-\frac{3\dddot{\alpha}}{t^{2}}+\frac{3\dot{\alpha}}{2t^{3}%
}+\frac{\Delta \alpha }{t^{3}}+\frac{2\Delta ^{2}\alpha }{t^{2}}\biggr\} 
\notag \\
&&-6\ddot{\phi}-\frac{12\dot{\phi}}{t}+\frac{4\Delta \phi }{t}+\frac{3\dot{%
\alpha}}{t}+\frac{2\Delta \alpha }{t}.  \label{z4}
\end{eqnarray}

\subsubsection{Large-scale limit of the pure quadratic theory ($B\rightarrow
\infty $) in the zero-shear gauge}

If we ignore the effects of matter\footnote{$\mu \propto a^{-3(1+\omega
)}\propto t^{-2}$ for radiation, whereas on the right-hand side of the
energy equation typical terms are $\sim \frac{\alpha }{t^{4}},\frac{\phi }{%
t^{4}}$}, i.e. $\delta T_{\nu }^{\mu }=0$, and take the large-scale limit
then each term in curly brackets vanishes. The combination $\frac{1}{t}(\ref%
{z2})+\frac{4}{3}(\ref{z1})-\frac{1}{t^{2}}(\ref{z3})$ gives 
\begin{equation}
2\frac{\dddot{\phi}}{t}+6\frac{\ddot{\phi}}{t^{2}}+\frac{3\dot{\phi}}{2t^{3}}%
=0,
\end{equation}%
to which the solution is

\begin{equation*}
\phi =C_{0}+C_{1}t^{-\frac{1}{2}}+C_{2}t^{\frac{1}{2}}.
\end{equation*}

Using the energy equation, we then obtain

\begin{equation*}
\alpha =C_{1}t^{-\frac{1}{2}}+3C_{2}t^{\frac{1}{2}}+C_{3}t^{\frac{3}{2}}.
\end{equation*}

So, at first sight it appears that there may be an instability as $%
t\rightarrow 0$; however, we note that the coefficients of the $t^{\frac{1}{2%
}}$ terms in $\alpha $ and $\phi $ are the same. This means that they can
both be locally defined away by the freedom to choose our initial time. If
we set $t=\tilde{t}-t_{0}$, then $a(t)=t^{\frac{1}{2}}=(\tilde{t}-t_{0})^{%
\frac{1}{2}}=\tilde{t}^{\frac{1}{2}}(1-t_{0}\tilde{t}^{-1})^{\frac{1}{2}%
}\sim \tilde{t}^{\frac{1}{2}}-\frac{1}{2}t_{0}\tilde{t}^{-\frac{1}{2}}$. In
fact, by choosing a different gauge, we see that this is just a curvature
perturbation when we compare our flat FRW background with the similar open
and closed models.

\subsection{Uniform Curvature Gauge ($\protect\delta R \equiv 0$)}

If we use the uniform curvature gauge then $R\equiv 0$ and we are perturbing
in flat space. The set of perturbations equations are now:

\noindent Energy constraint: 
\begin{eqnarray}
0 &=&-3\frac{\dot{\phi}}{t}+2\frac{\Delta \phi }{t}+\frac{3\alpha }{2t^{2}}-%
\frac{\Delta \chi }{t^{2}}-B\Biggl\{-3\frac{\dddot{\phi}}{t}-\frac{3\ddot{%
\phi}}{2t^{2}}+21\frac{\dot{\phi}}{t^{3}}+2\frac{\Delta \ddot{\phi}}{t}+6%
\frac{\Delta \dot{\phi}}{t^{2}}-7\frac{\Delta \phi }{t^{3}}  \notag \\
&&-2\frac{\Delta ^{2}\phi }{t^{2}}+\frac{3\ddot{\alpha}}{2t^{2}}-\frac{15%
\dot{\alpha}}{4t^{3}}-\frac{3\alpha }{2t^{4}}-\frac{7\Delta \alpha }{2t^{3}}-%
\frac{\Delta \ddot{\chi}}{t^{2}}+\frac{3\Delta \dot{\chi}}{2t^{3}}+\frac{%
11\Delta \chi }{2t^{4}}-\frac{\Delta ^{2}\chi }{t^{3}}\Biggr\}.  \label{s1}
\end{eqnarray}%
Momentum constraint: 
\begin{equation}
0=3\dot{\phi}-\frac{3\alpha }{2t}-B\Biggl\{3\dddot{\phi}+\frac{9\ddot{\phi}}{%
2t}-9\frac{\dot{\phi}}{t^{2}}-3\frac{\Delta \dot{\phi}}{t}+3\frac{\Delta
\phi }{t^{2}}-\frac{3\ddot{\alpha}}{2t}+\frac{9\dot{\alpha}}{4t^{2}}+\frac{%
3\Delta \alpha }{2t^{2}}-3\frac{\Delta \chi }{t^{3}}\Biggr\}.  \label{s2}
\end{equation}%
Trace-free propagation: 
\begin{equation}
0=-\phi -\alpha +\dot{\chi}+\frac{\chi }{2t}-B\Biggl\{-\ddot{\phi}-\frac{7%
\dot{\phi}}{2t}+\frac{\Delta \phi }{t}-\ddot{\alpha}+\frac{\dot{\alpha}}{2t}+%
\frac{\Delta \alpha }{t}+\dddot{\chi}+\frac{3\dot{\chi}}{4t^{2}}-\frac{3\chi 
}{4t^{3}}-\frac{\Delta \dot{\chi}}{t}-\frac{\Delta \chi }{2t^{2}}\Biggr\}.
\label{s3}
\end{equation}%
The trace equation is trivial, but we have a fourth equation from $\delta R=0
$: 
\begin{equation}
0=-3\ddot{\phi}-6\frac{\dot{\phi}}{t}+2\frac{\Delta {\phi }}{t}+\frac{3\dot{%
\alpha}}{2t}+\frac{\Delta \alpha }{t}-\frac{\Delta \dot{\chi}}{t}-\frac{%
\Delta \chi }{t^{2}}.  \label{s4}
\end{equation}

We can now eliminate $\chi $ from our equations (from the momentum
constraint, we see that $\chi $ is stable whenever $\alpha $ and $\phi $
are).

Using (\ref{s2}), equation (\ref{s4}) becomes: 
\begin{eqnarray}
0 &=&\frac{1}{B}\left[ t^{3}\ddot{\phi}+4t^{2}\dot{\phi}-\frac{t^{2}\dot{%
\alpha}}{2}-\frac{3t\alpha }{2}\right] -t^{3}\ddddot{\phi}-\frac{11}{2}t^{2}%
\dddot{\phi}-\frac{9}{2}t\ddot{\phi}  \label{s5} \\
&&+t^{2}\Delta \ddot{\phi}+2t\Delta \dot{\phi}+\frac{t^{2}\dddot{\alpha}}{2}+%
\frac{3t\ddot{\alpha}}{4}-\frac{t\Delta \dot{\alpha}}{2}.  \notag
\end{eqnarray}

Equation (\ref{s1}) becomes: 
\begin{eqnarray}
0 &=&-B\Biggl\{2\frac{\dddot{\phi}}{t}+3\frac{\ddot{\phi}}{t^{2}}-\Delta 
\dddot{\phi}+\frac{\Delta \ddot{\phi}}{2t}+5\frac{\Delta \dot{\phi}}{t^{2}}+%
\frac{\Delta ^{2}\dot{\phi}}{t}-3\frac{\Delta ^{2}{\phi }}{t^{2}}-\frac{%
\ddot{\alpha}}{t^{2}}+\frac{3\dot{\alpha}}{2t^{3}}-\frac{3\alpha }{2t^{4}}+%
\frac{\Delta \ddot{\alpha}}{2t}-\frac{7\Delta \dot{\alpha}}{4t^{2}}-\frac{%
\Delta ^{2}\alpha }{2t^{2}}\Biggr\}  \notag \\
&&-t\dddot{\phi}-\frac{3\ddot{\phi}}{2}+2\frac{\dot{\phi}}{t}+\frac{\Delta
\phi }{t}+\frac{\ddot{\alpha}}{2}-\frac{3\dot{\alpha}}{4t}+\frac{\alpha }{%
2t^{2}}+\frac{1}{B}\biggl[t\dot{\phi}-\frac{\alpha }{2}\biggr].  \label{s6}
\end{eqnarray}

\subsubsection{Large-scale limit of the pure quadratic theory ($B\rightarrow
\infty $) in the uniform curvature gauge}

In the large-scale limit of (\ref{s5}), setting $\phi =E_{\lambda
}t^{\lambda }$, $\alpha =F_{\lambda }t^{\lambda }$ and dropping the $B^{-1}$
terms, we find that

\begin{equation}
\lambda (\lambda -1)\left( \lambda -\frac{1}{2}\right) \left( (\lambda
+1)E_{\lambda }-\frac{F_{\lambda }}{2}\right) =0,
\end{equation}%
and in the large-scale limit of (\ref{s6}), keeping only the terms inside $%
\{\}$, we have 
\begin{equation}
(\lambda -1)\left( E_{\lambda }\lambda (2\lambda -1)-F_{\lambda }\left(
\lambda -\frac{3}{2}\right) \right) =0.
\end{equation}

The right-hand bracket of each of these last two equations cannot
simultaneously vanish unless $E_{\lambda }=F_{\lambda }=0$, so $\lambda =0,%
\frac{1}{2}$, or $1$ and $F_{0}=F_{\frac{1}{2}}=0$. There is a further
restriction on $E_{1}$ and $F_{1}$ from (\ref{s3}), specifically $F_1 = 
\frac{8}{5}E_1$ .

Hence, we have the solution 
\begin{eqnarray}
\phi  &=&E_{0}+E_{\frac{1}{2}}t^{\frac{1}{2}}+E_{1}t, \\
\alpha  &=&\frac{8}{5}E_{1}t, \\
\Delta \chi \equiv \Delta (t^{\frac{1}{2}}\beta +t\dot{\gamma}) &=&-\frac{3}{%
2}E_{\frac{1}{2}}t^{\frac{1}{2}}-\frac{9}{5}E_{1}t.
\end{eqnarray}%
Assuming $\Delta \chi =-\kappa ^{2}\chi $, we have 
\begin{eqnarray*}
\beta  &=&\frac{3}{2\kappa ^{2}}E_{\frac{1}{2}}-t^{\frac{1}{2}}\dot{g}(t,%
\mathbf{x}), \\
\gamma  &=&\frac{9}{5\kappa ^{2}}E_{1}t+g(t,\mathbf{x}).
\end{eqnarray*}

There is a remaining freedom in our gauge choice. In the conformal Newtonian
gauge, it is usually used to set both $\beta=0$ and $\gamma =0$. The other
parameters are unaffected by this choice. Here, we can use this freedom to
set $g(t, \mathbf{x})=0$.

Again, we see that the metric perturbations approach constant values as $%
t\rightarrow 0$ and the isotropic solution is stable against scalar
perturbations in this limit.

\section{Summary}

Combining our results from the quadratic theory, to leading order in the
time as $t\rightarrow 0$, we have found that the metric in the neighbourhood
of the special solution (\ref{n1}) with isotropic scale factor evolution $%
a=t^{1/2}$ takes the general form 
\begin{equation}
ds^{2}=-a^{2}(1+2\alpha )d\eta ^{2}-a^{2}\tilde{B}_{\alpha }d\eta dx^{\alpha
}+a^{2}(\delta _{\alpha \beta }+\tilde{C}_{\alpha \beta })dx^{\alpha
}dx^{\beta },  \label{metric}
\end{equation}%
where%
\begin{eqnarray}
\alpha  &=&\frac{8}{5}E_{1}t,  \label{f1} \\
\tilde{B}_{\alpha } &=&\frac{3}{\kappa ^{2}}E_{\frac{1}{2}\,,\alpha
}+2\left( D_{1}-\frac{2\Omega }{Bk^{4}}+2kD_{2}t^{\frac{1}{2}%
}+2k^{2}D_{1}t\right) Y_{\alpha }(\mathbf{x}),  \label{f2} \\
\tilde{C}_{\alpha \beta } &=&2\left( E_{0}+E_{\frac{1}{2}}t^{\frac{1}{2}%
}+E_{1}t\right) \delta _{\alpha \beta }+\frac{18t}{5\kappa ^{2}}%
E_{1\,,\alpha \beta }+  \label{f3} \\
&&+2\left( A_{1}+(2jA_{2}+A_{3})t^{\frac{1}{2}}+2j(jA_{1}+A_{4})t\right)
X_{\alpha \beta }(\mathbf{x}).
\end{eqnarray}%
where $\Delta Y_{\alpha }=-k^{2}Y_{\alpha }$, $\Delta X_{\alpha \beta
}=-j^{2}X_{\alpha \beta }$, the $A_{i}$ and $D_{i}$ are constants, and $%
E_{i}=E_{i}(\mathbf{x})$. 

These results are striking. The special isotropic and homogeneous solution
with scale factor evolution $a=t^{1/2}$ is found to be stable against the
effects of small scalar, vector and tensor perturbations in the $%
t\rightarrow 0$ limit. This adds to the evidence gathered from earlier
studies of the stability of this solution in the Bianchi type I, II and IX
universes \cite{bher1, cotIX}. It adds further weight to the conjecture that
small perturbations of this isotropic singularity form part of the general
cosmological solution for theories of gravity with quadratic lagrangians of
the sort studied here. This behaviour is quite different to that found in
cosmological solutions of general relativity, where isotropic expansion is
unstable as $t\rightarrow 0$. It arises because the higher-order quadratic
terms in the lagrangian contribute terms which dominate on approach to an
initial singularity (typically as $O(t^{-4})$ compared to general
relativistic contributions at $O(t^{-2})$) contribute isotropising stresses
which dominate over the shear stresses arising from expansion and
3-curvature anisotropy, which dominate the initial singularity in general
relativistic cosmologies.

These results have a number of implications for proposals to introduce
special cosmological initial conditions in order to ensure that the initial
state is isotropic, with low gravitational entropy. It appears to make
additional stipulations of special initial conditions unnecessary in
quadratic gravity. However, it may cause problems for the gravitational
entropy scenario in closed recollapsing universes because we expect the same
quadratic stresses also to drive the solution towards isotropy on approach
to any future 'big crunch' singularity. This would prevent it being the high
gravitational entropy state that a 'Second Law' of gravitational
thermodynamics would lead us to expect.

Finally, despite the special features of the cosmological models in
quadratic theories of gravity, we need to understand if these unusual
results regarding the stability of isotropy persist when
higher-than-quadratic-order contributions to the gravitational lagrangian
are included. In a subsequent paper we will present the results of this more
complicated study.

\textbf{Acknowledgements}: J. Middleton is supported by a PPARC studentship.
We would like to thank Sigbjorn Hervik and Timothy Clifton for discussions.

\newpage

\end{document}